\documentclass[twocolumn,secnumarabic,amssymb,amsmath,aps,prb,floatfix]{revtex4}
\usepackage{graphicx}
\usepackage{dcolumn}
\usepackage{bm}
\usepackage{amsmath}
\usepackage{braket}
\usepackage{array}
\usepackage{txfonts}

\newcommand\BaYbZnO{Ba$_3$Yb$_2$Zn$_5$O$_{11}$}

\begin{document}

\title{Low-Energy Excitations and Ground State Selection in Quantum Breathing Pyrochlore Antiferromagnet Ba$_3$Yb$_2$Zn$_5$O$_{11}$}
\author{T. Haku,$^1$ K. Kimura,$^2$ Y. Matsumoto,$^1$ M. Soda,$^1$ M. Sera,$^2$ 
D. Yu,$^3$ R. A. Mole,$^3$ T. Takeuchi,$^4$ 
S. Nakatsuji,$^1$ Y. Kono,$^1$ T. Sakakibara,$^1$ L.-J. Chang,$^5$ and 
T. Masuda$^1$}
\affiliation{
\\
$^1$ {\it Institute for Solid State Physics, The University of Tokyo, Kashiwa, Chiba 277-8581, Japan}
\\
$^2$ {\it Division of Materials Physics, Graduate School of Engineering Science, Osaka University, Toyonaka, Osaka 560-8531, Japan}
\\
$^3$ {\it Bragg Institute, Australian Nuclear Science and Technology Organization, Lucas Heights, New South Wales 2234, Australia}
\\
$^4${\it Low Temperature Center, Osaka University, Toyonaka, Osaka 560-0043, Japan}
\\
$^5${\it Department of Physics, National Cheng Kung University, Tainan 70101, Taiwan}
}
\date{\today}

\begin{abstract}
We study low energy excitations in the quantum breathing pyrochlore antiferromagnet \BaYbZnO\ 
by combination of inelastic neutron scattering (INS) and thermodynamical properties measurements. 
The INS spectra are quantitatively explained by spin-1/2 single-tetrahedron model having $XXZ$ 
anisotropy and Dzyaloshinskii-Moriya interaction. This model has a two-fold degeneracy of the 
lowest-energy state per tetrahedron and well reproduces the magnetization 
curve at 0.5 K and heat capacity above 1.5 K. At lower temperatures, however, 
we observe a broad maximum in the heat capacity around 63 mK, demonstrating that a unique 
quantum ground state is selected due to extra perturbations with energy scale smaller 
than the instrumental resolution of INS. 
\end{abstract}

\maketitle


In geometrically frustrated magnet, a macroscopic degeneracy remains in the ground state at zero 
temperature, as long as the geometry is preserved. 
Such a situation contradicts the third law of thermodynamics and small 
perturbations, which can induce non-trivial quantum states, play an important 
role in avoiding the breakdown of the basic law \cite{Geo_Rev,Balents}. 
A classic example of the violation of the third law is given by a regular tetrahedron 
of $S = 1/2$ Heisenberg spins; this has a nonmagnetic ground state with a two-fold degeneracy. 
In nature, however, neither perfect isolation nor absence of coupling to other degrees 
of freedom is achieved and a non-degenerate state is induced by a perturbation. 
In the presence of spin-lattice coupling the lifting of the degeneracy 
is accompanied by the distortion of the tetrahedron, which is called the spin 
Jahn-Teller effect \cite{SpinJT}. 
In the case of a three-dimensional (3D) lattice of corner 
sharing tetrahedra, i.e., the pyrochlore lattice\cite{SpinJT,Tchernyshyov}, 
the distortion is cooperatively propagated over the crystal, 
causing a magnetostructural phase transition\cite{ZnV2O4::1,ZnV2O4::2}. 
For an isolated regular tetrahedral system, on the other hand, 
experimental study is rare for lack of model compound. 
The search for a simple and isolated system 
is a challenge to the third law, leading to discovery of 
new state of matter at very low temperatures. 

In the absence of spin-lattice coupling in the Heisenberg spin pyrochlore system 
the degeneracy is lifted by 3D spin coupling of the magnetic ground state. 
This leads to quantum spin liquid\cite{pyro::theo::1}, 
ordering of spin singlet state\cite{pyro::theo::2,Berg03}, or chiral 
order state\cite{pyro::theo::3,pyro::theo::4}. 
The breathing pyrochlore lattice, i.e., one consisting of arrays 
of alternating large and small tetrahedra, has been found for 
the $S = 3/2$ spinels Li$A$Cr$_4$O$_8$ ($A$ = In and Ga)\cite{br::pyro::1,Goran}. 
The lattice is an 
experimental realization of a theoretical perturbation expansion 
method used for the pyrochlore lattice\cite{pyro::theo::1,pyro::theo::2,pyro::theo::3}. 
Theory predicts a spin 
liquid ground state for this model\cite{BrPy::theo}; 
however, LiInCr$_4$O$_8$\cite{Goran} exhibits a 
magnetostructural transition due to the spin Jahn-Teller effect 
similar to that observed in conventional pyrochlore  
compounds\cite{ZnV2O4::1,ZnV2O4::2,ACr2O4::1,ACr2O4::2,ACr2O4::3}. Thus the material 
that preserves the breathing pyrochlore geometry at low temperature will be important.

\BaYbZnO\ is an experimental realization of a breathing pyrochlore 
lattice formed by Yb$^{3+}$ ions\cite{br::pyro::2}, 
with both the small and large tetrahedra being regular. 
The oxygen ions surrounding the Yb$^{3+}$ ions are shared by 
the neighboring Yb$^{3+}$ ions in the small tetrahedra, 
while those are not shared in the large tetrahedra. 
This results in the small tetrahedra of Yb$_4$O$_{16}$ being surrounded by 
Zn$_{10}$O$_{20}$ supertetrahedra. 
This crystal structure suggests that intertetrahedra interaction is small 
and local distortion in the small tetrahedron, if it appears at low temperature, 
does not propagate to neighboring small tetrahedron. 
The magnetic susceptibility has been reported and can be explained by an $S$ = 1/2 
tetrahedron model; no phase transitions were observed with $T \ge {\rm 0.38}$ K\cite{br::pyro::2}. 
Crystalline electric field (CEF) excitations have been measured by inelastic neutron 
scattering (INS)\cite{BYZO::CEF}; 
the data were explained by four Kramers doublets with a first eigenenergy of 38.2 meV. 
This means that the low energy excitations are dominated by the ground state 
doublet and the effective spin 1/2 is a good approximation. 
Furthermore the eigenfunction of the ground state was shown to exhibit an easy-plane 
type magnetic moment. Even including this anisotropy term the ground state 
of the tetrahedral spin system is a doublet in the absence of any 
intertetrahedron interaction or spin-lattice coupling. 
As such \BaYbZnO\ is a candidate for the classic example of frustrated magnets. 
In this communication we study low energy excitations to identify the effective spin Hamiltonian by INS experiment and macroscopic properties at very low temperatures to see how nature keeps the third law of thermodynamics.
We demonstrate how the degeneracy of the ground state is lifted and unique quantum 
state is selected in \BaYbZnO .


INS experiments were performed
using the neutron spectrometer 
PELICAN\cite{PELICAN_Perf1} at ANSTO. 
We utilized setup I using an incident energy $E_i$ of 2.1 meV 
and setup II using an $E_i$ of 3.6 meV. 
Setup I afforded a resolution of 0.059 meV full width half maximum (FWHM) at 
the elastic line, while setup II gave 0.135 meV. 


\begin{figure}[htbp]
\begin{center}
\includegraphics[width=85mm]{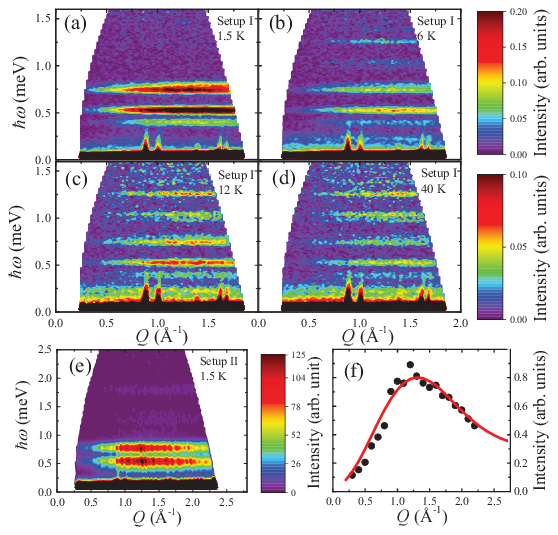}
\caption{\label{fig1}
(a)- (d) INS spectra measured at 1.5 K (a), 6 K (b), 12 K (c), and 40 K (d) using 
setup I.
(e) INS spectrum measured at 1.5 K using the setup II. 
(f) $Q$ dependence of the integrated intensity obtained from spectrum in (e). 
Red solid curve is the calculation (see text). 
}
\end{center}
\end{figure}

\begin{figure}[htbp]
\begin{center}
\includegraphics[width=85mm]{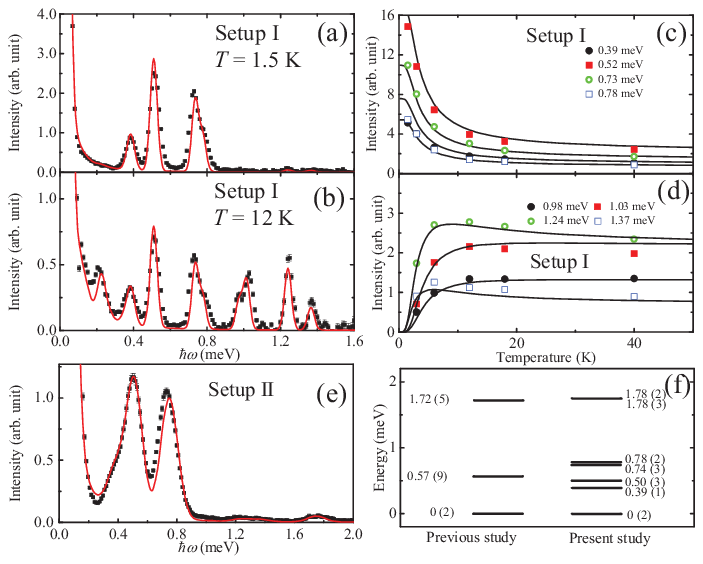}
\caption{\label{fig2}
(a), (b) The $\hbar \omega$ dependences of the neutron intensities 
at 1.5 K for (a) and 12 K for (b). 
(c) Temperature dependences of the intensities of the excitations in (a). 
(d) Those of the excitations additionally observed in high temperatures in (b). 
(e) $\hbar \omega$ dependence of the intensity at 1.5 K obtained using setup II. 
Throughout the panels red and black solid curves are the calculation using the same parameters 
in Eqs.~(3)-(6). 
(f) Energy level of $S = 1/2$ Heisenberg spin tetrahedron model in the 
previous study\cite{br::pyro::2}  and that of $S = 1/2$ anisotropic 
spin tetrahedron in the present study. 
}
\end{center}
\end{figure}

\begin{figure}[htbp]
\begin{center}
\includegraphics[width=85mm]{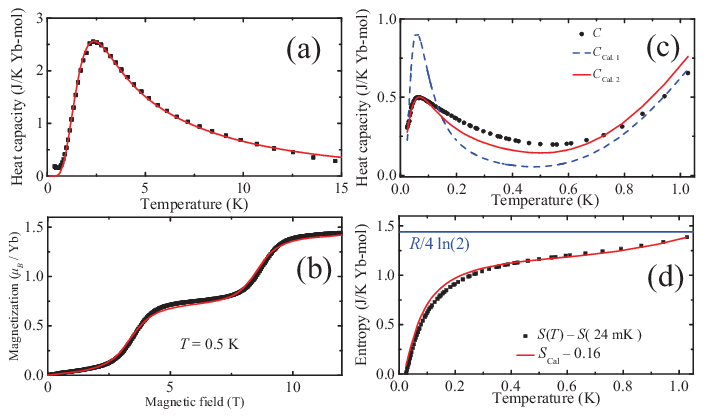}
\caption{\label{fig4}
(a) Heat capacity measured 
reported in the previous study\cite{br::pyro::2}. 
Red solid curve is calculation. 
(b) Magnetic field dependence of magnetization at 0.5 K. Red solid curve is 
calculation. 
(c) Filled circles indicate heat capacity. 
Blue dashed curve is the calculation of the empirical model where 
the lift of the degeneracy of the ground state is introduced as a single energy gap. 
Red solid curve is the calculation of the model where the empirical energy gap has 
distribution. 
(d) Entropy change estimated from (c). The calculated entropy 
is shifted so that the calculation has the same value of the data at 0.94 K. 
}
\end{center}
\end{figure}

The INS spectra measured using setup I are 
shown in Figs. \ref{fig1}(a)-\ref{fig1}(d). 
Three flat bands are observed at 1.5 K; the absence of dispersion suggests 
that these bands are approximately cluster excitations and the effect of 
any intercluster excitation is small and hidden in the instrument resolution. 
At 6 K the intensities of these three excitations are suppressed and additional 
flat bands are observed at different $\hbar \omega$'s. 
In all panels there are several 
streaks observed in the range $\hbar \omega \lesssim$ 0.4 meV 
which was ascribed to acoustic phonons. 
Figure \ref{fig1}(e) shows the INS spectrum obtained using setup II 
while panel (f) shows the $Q$ dependence of the intensity integrated 
over the range $0.25~{\rm meV} < \hbar \omega < 0.95~{\rm meV}$.

The symbols in Figs. \ref{fig2}(a) and \ref{fig2}(b) shows the one dimensional energy 
cuts from the data presented in Figs.~\ref{fig1}(a) and \ref{fig1}(c), respectively. 
The peaks are fitted by Gaussian functions with FWHM restricted to that of instrumental resolution 
to estimate the peak energies and the intensities. 
The peak energies at 1.5 K are 0.39, 0.52, 0.73 and 0.78 meV. At 12 K additional peaks 
are observed
in Fig.~\ref{fig2}(b). 
The temperature dependence of the four excitations observed at 1.5 K are shown in Fig.~\ref{fig2}(c), 
while the additional excitations at 12 K are shown in Fig.~\ref{fig2}(d). 
The former monotonically decrease with increasing temperature 
while the latter show the opposite behavior. 
This implies that those excitations at 1.5 K are ground 
state transitions and those at 12 K originate from excited states. 

Figure \ref{fig2}(e) shows one-dimensional energy cuts from the INS spectra at 1.5 K in Fig.~\ref{fig1}(e) 
using setup II. 
A peak is observed at 1.75 meV in addition to the peaks in setup I. 
The $Q$ dependence of the intensity 
in Fig.~\ref{fig1}(f) exhibits broad maximum at $Q_{\rm max} \sim 1.25 {\rm \AA}^{-1}$. 
This means that antiferromagnetic correlation between the spins, the characteristic length scale of which 
is ${\pi}/Q_{\rm max}$, is enhanced. 
The dispersionless excitations with the $Q$ dependent intensity means that the 
neutron spectrum is dominated by an antiferromagnetic cluster within the instrumental resolution. 

For analysis on INS spectra we assume a spin tetrahedron model. 
The number of the excitations observed at the base temperature is four, which is inconsistent 
with Heisenberg $S$ = 1/2 spin tetrahedron. 
We, therefore, consider the following general expression\cite{SL}: 
\begin{align}
\mathcal{H} = - \sum_{i<j} \sum_{\nu\mu}
J_{ij}^{\nu\mu} S_i^{\nu} S_j^{\mu}. 
\label{hamiltonian}
\end{align}
Here $i$ and $j$ are the labels of the spins on the tetrahedron 
and ${\nu}$ and ${\mu}$ represent Cartesian coordinates $x$, $y$, and $z$ 
which are defined along the crystallographic $a$, $b$ and $c$-axes, respectively. 
The position vectors for the spins are
${\bm r}_1 = d/\sqrt{3}(1,1,1)$,
${\bm r}_2 = d/\sqrt{3}(-1,-1,1)$,
${\bm r}_3 = d/\sqrt{3}(1,-1,-1)$, and 
${\bm r}_4 = d/\sqrt{3}(-1,1,-1)$, 
where $2{\sqrt 2}d$ is the length of the side of the tetrahedron. 
The symmetry of the regular tetrahedron determines 
the form of the interaction tensor $\hat{J}_{ij}$\cite{SL}, 
for example in the case of $\hat{J}_{12}$ one gets, 
\begin{align}
\hat{J}_{12} = 
\begin{pmatrix}
J_{xx} & J_{yx}& J_{zx} \\
J_{xy} & J_{yy}& J_{zy} \\
J_{xz} & J_{yz}& J_{zz} 
\end{pmatrix}
=
\begin{pmatrix}
J_{1} & J_{3}& -J_{4} \\
J_{3} & J_{1}& -J_{4} \\
J_{4} & J_{4}& J_{2} 
\end{pmatrix}
\label{interaction}.
\end{align}
Using the Hamiltonian in Eq. (\ref{hamiltonian}) and the interaction tensor $\hat{J}_{ij}$, 
we calculate the neutron scattering cross section. 
We performed a numerical fitting to the peak energies and their intensities measured 
at $T$ = 1.5 K in setup I using a genetic algorithm. 
The fitting parameters are $J_1$, $J_2$, $J_3$, $J_4$ and a scale factor. 
Here $J_1$ is an $XY$ type interaction, $J_2$ is Ising type, $J_3$ 
is a pseudo-dipole-type and $J_4$ is Dzyaloshinskii-Moriya (DM) type. 
The magnitude of the Dzyaloshinskii correspond to $\sqrt{2}J_4$. 
The best fit parameters are:
\begin{align}
J_1 = -0.570 \pm 0.033 ~{\rm meV},\label{parameters1}
\\
J_2 = -0.558 \pm 0.028 ~{\rm meV},\label{parameters2}
\\
J_3 = 0.000 \pm 0.023 ~{\rm meV},\label{parameters3}
\\
J_4 = 0.113 \pm 0.014 ~{\rm meV}.\label{parameters4}
\end{align}
These are consistent within error to those obtained from 
independent INS experiments reported recently by another group\cite{recent}. 
The results show 
that the system 
has an easy-plane $XXZ$ type anisotropy with the DM-type interaction about 30 \% 
of the main interaction. The predicted intensities have been convoluted with 
the instrumental resolution and are indicated by the solid red curves 
in Figs.~\ref{fig2}(a) and \ref{fig2}(b). 
Similarly the predicted temperature dependence is shown in Figs.~\ref{fig2}(c) and \ref{fig2}(d) 
by the solid black curves. All the data are reproduced by the calculations. 
The data collected using setup II in Fig.~\ref{fig1}(e) 
is also reproduced using the same set of parameters. 
Thus we can conclude that the neutron data are explained by $S = 1/2$ spin 
tetrahedron including DM interaction in the temperature range $1.5~{\rm K} \le T \le 40~{\rm K}$.

The energy levels obtained in the INS experiment are shown in right hand panel of 
Fig.~\ref{fig2}(f). The eigenenergies are calculated using the parameters 
from Eqs.~(\ref{parameters1})-(\ref{parameters4}). 
For reference the energy level obtained by bulk measurements\cite{br::pyro::2} 
on the basis of 
Heisenberg model is shown in the left panel. 
The energy levels of the Heisenberg model is characterized by a total spin 
${\bm S_{\rm total}}= {\bm S_1}+{\bm S_2}+{\bm S_3}+{\bm S_4}$. 
The nine-fold degeneracy of the first excited state of $S$ = 1 
is lifted by the $XXZ$ anisotropy and DM interactions. 
Similarly the degeneracy of the $S$ = 2 state is lifted. 
The doublet of the ground state is, however, lifted neither by the anisotropy nor DM interaction in the framework of 
isolated tetrahedron Hamiltonian. 
In the present model, total spin $S_{\rm total}$ is no longer a good quantum number because 
DM interaction mixes the eigenstates in isotropic Hamiltonian. 
This modifies the selection rules of neutron scattering and allows finite matrix components for 
the transitions among all the eigenstates 
and leads to the observation of many excitations in the INS spectra.

The anisotropic exchange parameters determined using INS was then 
used to calculate the thermodynamic properties of the 
Hamiltonian Eq.~(\ref{hamiltonian}) and a comparison to experiment was made. 
Figure \ref{fig4}(a) shows the heat capacity calculated using Eq. (\ref{hamiltonian}), 
together with that measured in a previous study\cite{br::pyro::2}. 
The data is reasonably reproduced by the calculation for 
$T \ge$ 1.5 K; a broad peak at $T \sim$ 2.5 K indicates the Schottky anomaly 
associated with the excited states in the range 0.39 meV $\le \hbar \omega \le$ 0.78 meV 
as shown in Fig. \ref{fig2}(f). 
An excellent agreement between experiment and calculation is also observed for 
the full magnetization curve at 0.5 K in Fig. \ref{fig4}(b). 
Here the best reproduction of the data was with an anisotropic $g$ 
tensors $g_{\perp}$ = 2.78 and $g_{\parallel}$ = 2.22, where the principle axis 
of the $g$ factor is taken along $\hat{r}_i$. 
This easy-plane anisotropy of the $g$-tensor is consistent with the previous study. 
The magnetization curve exhibits two pronounced steps at 
$H_{C1} \sim 3.5$ T and $H_{C2} \sim 8.8$ T; these correspond to the transitions from 
the doublet ground state to the first set of excited states in the range of 
0.39 meV $\le \hbar \omega \le $0.78 meV and the first set to the second set at 
$\hbar \omega \sim$ 1.75 meV in Fig.~\ref{fig2}(f). 
The effect of the anisotropic exchange interactions manifests in the non-equivalent 
distance of the steps and the ramp-like structure rather than the stair-like structure. 
The former are not expected for a Heisenberg spin tetrahedron model. 
The anisotropic exchange interactions mix the $S_{\rm total}$ = 0 and $S_{\rm total}$ = 1 
states, giving rise to a finite magnetic moment of the lowest energy doublet with 
$\braket{S_{\rm total}}_{\rm GS}= 0.13$, 
which is consistent with the finite slope at $H < 2.5$ T. 

Thus, the anisotropic $S = 1/2$ single tetrahedral model can account for multiple 
experimental data: the INS spectra at 1.5 K in the range of $\hbar \omega \gtrsim 0.15$ meV, 
the magnetization curve at 0.5 K, and heat capacity above 1.5 K. 
However, this model has a two-fold degeneracy of the lowest energy state, from which the 
real ground state should be selected by additional interactions 
that are not included in this model. 
Therefore we performed heat capacity measurements in the range 24 mK to 1 K, 
the results of this are shown in Fig. \ref{fig4}(c). 
Rather than a sharp peak indicative of a phase transition, 
the heat capacity exhibits a broad peak at $T \sim$ 63 mK. 
The entropy change per Yb$^{3+}$ ion is calculated to be 1.4 J/K $ \sim R/4 \ln(2)$; 
the calculation is shown as the solid line in Fig. \ref{fig4}(d). 
Such a change in entropy corresponds to the release of two degrees 
of freedom per spin-tetrahedron. Thus, the heat capacity measurements 
at low temperature demonstrate that a unique ground state is finally 
selected from the doublet ground state of the single tetrahedral model.

To explain the heat capacity and entropy change, 
we firstly assume that the doublet ground state of all the spin tetrahedra 
is lifted by a single energy gap $E_g$. 
This corresponds to the assumption that all the Yb$_4$ tetrahedra are uniformly distorted. 
The dashed curve in Fig.~\ref{fig4}(c) is the calculation using $E_g$ = 0.012 meV; 
the calculated heat capacity is dominated by Schottky behavior from 
the two level system of the split doublet. 
The peak is much narrower than that of experiment and the model does not 
explain the data. Secondly, we assume that $E_g$ has distribution, which includes 
the possibilities of different $E_g$ for each spin tetrahedron and dispersive $E_g$ 
due to inter-tetrahedra interaction. We use Lorentzian function for the distribution 
of $E_g$ with the peak center $E_c$ = 0.010 meV and the FWHM $E_l$ = 0.016 meV. 
The calculated heat capacity and entropy change are 
indicated by red solid curves. 
The calculation reasonably reproduces the data. 
The doublet ground state of the isolated tetrahedral 
Hamiltonian identified within the instrumental resolution of 
INS experiment is, thus, lifted by a perturbation. 
Furthermore the energy gap exhibits distribution or dispersion. 

The ground state of \BaYbZnO\ is, therefore, not a solution of the spin 
Hamiltonian in Eq.~(\ref{hamiltonian}) but a non-trivial quantum state. 
The possible perturbations for lifting this degeneracy should be discussed; 
one possibility is that there is an interaction between spin tetrahedra, 
although this is small. The theory of Heisenberg spin tetrahedra 
predicts that a partial ordering is induced and the energy gap is estimated as 
$10^{-3}J_{\rm inter}^3/ 48 J_{\rm intra}^2$\cite{pyro::theo::2}, 
where $J_{\rm intra}$ and $J_{\rm inter}$ are the intra and inter tetrahedron interactions, respectively. 
Since no dispersion is observed within the instrumental resolution of 0.059 meV 
we can assume that the band width of the excitations is smaller than this value. 
This leads to the estimation of an upper bound for $J_{\rm inter}$ of 0.015 meV 
using the RPA approximation. 
Note that $J_{\rm intra} \sim$ 0.5 meV in Eq. (\ref{parameters1}), 
and the magnitude of the calculated gap is too small to explain the observed gap. 
However a large asymmetric interaction, $J_4$,  
is obtained, this is 20\% of $J_1$ and $J_2$. 
The theory considering DM interactions with a magnitude of a few percent of the main 
Heisenberg interaction suggests that the energy gap of the partial ordering of dimers is 
qualitatively enhanced\cite{DM_pyro}. 
Furthermore a chiral ordered phase is predicted at lower temperatures. 
The candidates for the quantum ground state 
are therefore either a partial ordering of dimers or a chiral ordered 
state induced by a combination of inter-tetrahedra interactions and a large DM interaction. 

Another possibility for the perturbation is spin lattice coupling. 
In a single spin $S = 1/2$ tetrahedron coupled to the lattice, the ground state doublet is 
lifted by the spin Jahn-Teller (JT) mechanism\cite{SpinJT}. 
In the spinel compounds ZnV$_2$O$_4$ and MgV$_2$O$_4$\cite{SpinJT,Tchernyshyov} 
a magnetostructural transitions was observed, which was due to a coupling 
of the interaction pathways and the three dimensional lattice. 
While in MgCr$_2$O$_4$ a precursor to the spin JT was observed above 
the transition due to the spin dynamical JT (DJT) effect\cite{Watanabe}. 
In contrast to the uniform array of tetrahedra in the pyrochlore lattice, the Yb$_4$O$_{16}$ 
tetrahedra in \BaYbZnO\ are surrounded by JT inactive Zn$_{10}$O$_{20}$ 
supertetradra and are comparatively isolated. 
This circumstance is quite similar to that in the honeycomb compound 
Ba$_3$CuSb$_2$O$_9$\cite{Zhou,Nakatsuji}, where the Cu$^{2+}$ 
tetrahedra are face-shared by JT-inactive SbO$_6$ octahedra and thus isolated; 
this suppresses the static JT distortion and a quantum spin liquid is induced 
by DJT\cite{Nasu,Han}. In analogy to Ba$_3$CuSb$_2$O$_9$ 
the spin DJT is a candidate for the microscopic mechanism of the suppression 
of the structural transition and the appearance of a quantum spin liquid in 
\BaYbZnO . However, no equivalent theory has been reported for the 
breathing pyrochlore lattice and the ground state is still an open question.


\begin{acknowledgments}
Prof. Tsuyoshi Kimura is greatly appreciated for helpful discussion. 
T. Haku was supported by the Japan Society for the Promotion of Science 
through the Program for Leading Graduate Schools (MERIT).
This work was supported by JSPS KAKENHI Grant in Aid for 
Scientific Research (B) Grant No. 24340077 and 24340075.
Travel expenses for the experiment performed using 
PELICAN at ANSTO, Australia, were supported by General User Program for 
Neutron Scattering Experiments, Institute for Solid State Physics, 
The University of Tokyo (proposal No. 15543), at JRR-3, 
Japan Atomic Energy Agency, Tokai, Japan.
Magnetization measurement was carried out by the joint research in 
the Institute for Solid State Physics, the University of Tokyo. 
\end{acknowledgments}

\providecommand{\noopsort}[1]{}\providecommand{\singleletter}[1]{#1}%


\begin{thebibliography}{99}
\providecommand*{\bibinfo}[2]{#2}
\providecommand*{\eprint}[1]{#1}
\providecommand*{\url}[1]{#1}
\bibitem{Geo_Rev}
\bibinfo{author}{A.~P. Ramirez}, \bibinfo{journal}{Annu. Rev. Mater. Sci.}
\bibinfo{volume}{\textbf{24}}, \bibinfo{pages}{453} (\bibinfo{date}{1994}).
\bibitem{Balents} L. Balents, Nature {\bf 464}, 199 (2010). 
\bibitem{SpinJT}
\bibinfo{author}{Y.~Yamashita} and \bibinfo{author}{K.~Ueda},
\bibinfo{journal}{Phys. Rev. Lett.} \bibinfo{volume}{\textbf{85}},
\bibinfo{pages}{4960} (\bibinfo{date}{2000}).
\bibitem{Tchernyshyov}
O. Tchernyshyov, R. Moessner, and S. L. Sondhi, 
Phys. Rev. Lett. {\bf 88}, 067203 (2002). 
\bibitem{ZnV2O4::1}
\bibinfo{author}{Y.~Ueda}, \bibinfo{author}{N.~Fujiwara}, and
\bibinfo{author}{H.~Yasuoka}, \bibinfo{journal}{Journal of the Physical
Society of Japan} \bibinfo{volume}{\textbf{66}}(3), \bibinfo{pages}{778}
(\bibinfo{date}{1997}).
\bibitem{ZnV2O4::2}
\bibinfo{author}{H.~Mamiya} and \bibinfo{author}{M.~Ononda},
\bibinfo{journal}{Solid State Commun.} \bibinfo{volume}{\textbf{95}},
\bibinfo{pages}{217} (\bibinfo{date}{1995}).
\bibitem{pyro::theo::1}
\bibinfo{author}{B.~Canals} and \bibinfo{author}{C.~Lacroix},
\bibinfo{journal}{Phys. Rev. Lett.} \bibinfo{volume}{\textbf{80}},
\bibinfo{pages}{2933} (\bibinfo{date}{1998}).
\bibitem{pyro::theo::2}
\bibinfo{author}{H.~Tsunetsugu}, \bibinfo{journal}{Phys. Rev. B}
\bibinfo{volume}{\textbf{65}}, \bibinfo{pages}{024415}
(\bibinfo{date}{2001}).
\bibitem{Berg03}
E. Berg, E. Altman, and A. Auerbach, Phys. Rev. Lett. {\bf 90}, 147204 (2003). 
\bibitem{pyro::theo::3}
\bibinfo{author}{J.~H. Kim} and \bibinfo{author}{J.~H. Han},
\bibinfo{journal}{Phys. Rev. B} \bibinfo{volume}{\textbf{78}},
\bibinfo{pages}{180410(R)} (\bibinfo{date}{2008}).
\bibitem{pyro::theo::4}
\bibinfo{author}{F.~J. Burnell}, \bibinfo{author}{S.~Chakravarty}, and
\bibinfo{author}{S.~L. Sondhi}, \bibinfo{journal}{Phys. Rev. B}
\bibinfo{volume}{\textbf{79}}, \bibinfo{pages}{144432}
(\bibinfo{date}{2009}).
\bibitem{br::pyro::1}
\bibinfo{author}{Y.~Okamoto}, \bibinfo{author}{G.~J. Nilsen},
\bibinfo{author}{J.~P. Attfield}, and \bibinfo{author}{Z.~Hiroi},
\bibinfo{journal}{Phys. Rev. Lett.} \bibinfo{volume}{\textbf{110}},
\bibinfo{pages}{097203} (\bibinfo{date}{2013}).
\bibitem{Goran} 
G. J. Nilsen, Yoshihiko Okamoto, Takatsugu Masuda, Juan Rodriguez-Carvajal, Hannu Mutka, 
Thomas Hansen, and Zenji Hiroi, 
Phys. Rev. B {\bf 91}, 174435 (2015). 
\bibitem{BrPy::theo}
\bibinfo{author}{O.~Benton} and \bibinfo{author}{N.~Shannon},
\bibinfo{journal}{J. Phys. Soc. Jpn} \bibinfo{volume}{\textbf{84}},
\bibinfo{pages}{104710} (\bibinfo{date}{2015}).
\bibitem{ACr2O4::1}
\bibinfo{author}{S.~H.~Lee}, \bibinfo{author}{C.~Broholm}, \bibinfo{author}{T.~H.
Kim}, \bibinfo{author}{W.~Ratcliff}, and \bibinfo{author}{S.-W. Cheong},
\bibinfo{journal}{Phys. Rev. Lett.} \bibinfo{volume}{\textbf{84}},
\bibinfo{pages}{3718} (\bibinfo{date}{2000}).
\bibitem{ACr2O4::2}
\bibinfo{author}{M.~Matsuda}, \bibinfo{author}{H.~Ueda},
\bibinfo{author}{A.~Kikkawa}, \bibinfo{author}{Y.~Tanaka},
\bibinfo{author}{K.~Katsumata}, \bibinfo{author}{Y.~Narumi},
\bibinfo{author}{T.~Inami}, \bibinfo{author}{Y.Ueda}, and
\bibinfo{author}{S.-H. Lee}, \bibinfo{journal}{Nat. Phys.}
\bibinfo{volume}{\textbf{3}}, \bibinfo{pages}{397} (\bibinfo{date}{2007}).
\bibitem{ACr2O4::3}
\bibinfo{author}{H.~Ueda}, \bibinfo{author}{H.~A. Katori},
\bibinfo{author}{H.~Mitamura}, \bibinfo{author}{T.~Goto}, and
\bibinfo{author}{H.Takagi}, \bibinfo{journal}{Phys. Rev. Lett.}
\bibinfo{volume}{\textbf{94}}, \bibinfo{pages}{047202}
(\bibinfo{date}{2005}).
\bibitem{br::pyro::2}
\bibinfo{author}{K.~Kimura}, \bibinfo{author}{S.~Nakatsuji}, and
\bibinfo{author}{T.~Kimura}, \bibinfo{journal}{Phys. Rev. B}
\bibinfo{volume}{\textbf{90}}, \bibinfo{pages}{060414(R)}
(\bibinfo{date}{2014}).
\bibitem{BYZO::CEF}
\bibinfo{author}{T.~Haku}, \bibinfo{author}{M.~Soda},
\bibinfo{author}{M.~Sera}, \bibinfo{author}{K.~Kimura},
\bibinfo{author}{S.~Itoh}, \bibinfo{author}{T.~Yokoo}, and
\bibinfo{author}{T.~Masuda}, \bibinfo{journal}{Journal of the Physical
Society of Japan} \bibinfo{volume}{\textbf{85}}, \bibinfo{pages}{034721}
(\bibinfo{date}{2016}).
\bibitem{PELICAN_Perf1}
\bibinfo{author}{D.~Yu}, \bibinfo{author}{R.~Mole},
\bibinfo{author}{T.~Noakes}, \bibinfo{author}{S.~Kennedy}, and
\bibinfo{author}{R.~Robinson}, \bibinfo{journal}{Journal of the Physical
Society of Japan} \bibinfo{volume}{\textbf{82}}, \bibinfo{pages}{SA027}
(\bibinfo{date}{2013}).
\bibitem{Sakakibara}
T. Sakakibara, H. Mitamura, T. Tayama and H. Amitsuka, Jpn. J. Appl. Phys. {\bf 33}, 5067 (1994). 
\bibitem{SL}
\bibinfo{author}{K.~A. Ross}, \bibinfo{author}{L.~Savary},
\bibinfo{author}{B.~D.~Gaulin}, and \bibinfo{author}{L.~Balents},
\bibinfo{journal}{Phys. Rev. X} \bibinfo{volume}{\textbf{1}},
\bibinfo{pages}{021002} (\bibinfo{date}{2011}).
\bibitem{recent}
J. G. Rau, L. S. Wu, A. F. May, L. Poudel, B. Winn, V. O. Garlea, A. Huq, 
P. Whitfield, A. E. Taylor, M. D. Lumsden, M. J. P. Gingras, and A. D. Christianson, 
arxiv:1601.04104.
\bibitem{Squires}
\bibinfo{author}{G.~Squires}, \bibinfo{title}{\emph{Introduction to the theory
of THERMAL NEUTRON SCATTERING}} (\bibinfo{publisher}{Cambridge University
Press}, Cambridge, \bibinfo{year}{2002}).
\bibitem{iden}
\bibinfo{author}{J.~Holland}, \bibinfo{title}{\emph{Adaptation in Natural and
Artifical Systems}} (\bibinfo{publisher}{University of Michigan Press}, MI,
\bibinfo{year}{1975}).
\bibitem{DM_pyro}
\bibinfo{author}{V.~N. Kotov}, \bibinfo{author}{M.~Elhajal},
\bibinfo{author}{M.~E. Zhitomirsky}, and \bibinfo{author}{F.~Mila},
\bibinfo{journal}{Phys. Rev. B} \bibinfo{volume}{\textbf{72}},
\bibinfo{pages}{014421} (\bibinfo{date}{2005}).
\bibitem{Watanabe}
T. Watanabe, S-I. Ishikawa, H. Suzuki, Y. Kousaka, and K. Tomiyasu, 
Phys. Rev. B {\bf 86}, 144413 (2012). 
\bibitem{Zhou}
H. D. Zhou, E.S. Choi, G. Li, L. Balicas, C.R. Wiebe, Y. Qiu, J.R.D. Copley, and J.S. Gardner, 
Phys. Rev. Lett. {\bf 106}, 147204 (2011). 
\bibitem{Nakatsuji}
S. Nakatsuji, K. Kuga, K. Kimura, R. Satake, N. Katayama, 
E. Nishibori, H. Sawa, R. Ishii, M. Hagiwara, F. Bridges. T. U. Ito, 
W. Higemoto, Y. Karaki, M. Halim, A. A. Nugroho, J. A. Rodriguez-Rivera, M. A. Green, 
and C. Broholm, Science {\bf 336}, 559 (2012). 
\bibitem{Nasu}
J. Nasu and S. Ishihara, Phys. Rev. B {\bf 88}, 094408 (2013). 
\bibitem{Han}
Y. Han, M. Hagiwara, T. Nakano, Y. Nozue, K. Kimura, M. Halim, and S. Nakatsuji, 
Phys. Rev. B {\bf 92}, 180410(R) (2015). 

\end{thebibliography}
\end{document}